\documentclass{pasj00}
\draft

\begin{document}
\SetRunningHead{Bamba et al.}{HESS J1804$-$216 with Suzaku}
\Received{2006/07/15}
\Accepted{2006/08/14}

\title{Discovery of a possible X-ray counterpart to HESS~J1804$-$216}

\author{Aya \textsc{Bamba}\altaffilmark{1},
Katsuji \textsc{Koyama}\altaffilmark{2},
Junko S. \textsc{Hiraga}\altaffilmark{1},
John P. \textsc{Hughes}\altaffilmark{3},
Takayoshi \textsc{Kohmura}\altaffilmark{4},
\\
Motohide \textsc{Kokubun}\altaffilmark{5},
Yoshitomo \textsc{Maeda}\altaffilmark{6},
Hironori \textsc{Matsumoto}\altaffilmark{2},
Atsushi \textsc{Senda}\altaffilmark{1},
\\
Tadayuki \textsc{Takahashi}\altaffilmark{6},
Yohko \textsc{Tsuboi}\altaffilmark{7},
Shigeo \textsc{Yamauchi}\altaffilmark{8},
Takayuki \textsc{Yuasa}\altaffilmark{5}
}
\altaffiltext{1}{RIKEN, cosmic radiation group,
2-1, Hirosawa, Wako-shi, Saitama, Japan}
\email{bamba@crab.riken.jp}
\altaffiltext{2}{Department of Physics, Graduate School of Science,\\
Kyoto University, Sakyo-ku, Kyoto 606-8502, Japan}
\altaffiltext{3}{Department of Physics and Astronomy, Rutgers University,\\
136 Frelinghuysen Road, Piscataway, NJ 08854-8019, USA}
\altaffiltext{4}{kougakuin University,
2665-1, Nakano-cho, Hachioji, Tokyo 192-0015, Japan}
\altaffiltext{5}{Department of Physics, Graduate School of Science,
University of Tokyo,\\
Hongo 7-3-1, Bunkyo-ku, Tokyo 113-0033, Japan}
\altaffiltext{6}{Department of High Energy Astrophysics,
Institute of Space and Astronautical Science (ISAS),\\
Japan Aerospace Exploration Agency (JAXA),\\
3-1-1 Yoshinodai, Sagamihara, Kanagawa 229-8510, Japan}
\altaffiltext{7}{Department of Physics, Faculty of Science and Engineering,
Chuo University,\\
1-13-27 Kasuga, Bunkyo-ku, Tokyo 112-8551, Japan}
\altaffiltext{8}{Faculty of Humanities and Social Sciences,
Iwate University, \\
3-18-34 Ueda, Morioka, Iwate 020-8550, Japan}


%

\KeyWords{acceleration of particles
--- ISM: individual (HESS~J1804$-$216)
--- X-rays: ISM
--- X-rays: individual (Suzaku~J1804$-$2142, Suzaku~J1804$-$2140)
} 

\maketitle

\begin{abstract}
Suzaku deep observations have discovered
two highly significant nonthermal X-ray sources,
Suzaku~J1804$-$2142 (Src~1) and Suzaku~J1804$-$2140 (Src~2),
positionally coincident with the unidentified 
TeV $\gamma$-ray source, HESS~J1804$-$216.
The X-ray sources are not time variable and show no counterpart
in other wavebands, except for the TeV source.
Src~1 is unresolved at Suzaku spatial resolution,
whereas Src~2 is extended or composed of multiple sources.
The X-ray spectra are highly absorbed, hard, and featureless,
and are well fitted by absorbed power-law models with 
best-fit photon indices and absorption columns of 
$-0.3_{-0.5}^{+0.5}$ and $0.2_{-0.2}^{+2.0}\times 
10^{22}$~cm$^{-2}$ for Src~1,
and
$1.7_{-1.0}^{+1.4}$ and $1.1_{-0.6}^{+1.0}\times 
10^{23}$~cm$^{-2}$ for Src~2.
The measured X-ray absorption to the latter
source is significantly larger than the total Galactic neutral hydrogen
column in that direction.
The unabsorbed 2--10~keV band luminosities are
$7.5\times 10^{32}(d/{\rm 5~kpc})^2$~ergs~s$^{-1}$ (Src 1) and
$1.3\times 10^{33}(d/{\rm 5~kpc})^2$~ergs~s$^{-1}$ (Src 2),
where $d$ is the source distance.
Among the handful of TeV sources with known X-ray counterparts, 
HESS J1804$-$216 has the largest ratio of TeV $\gamma$-ray 
to hard X-ray fluxes. We discuss the nature of the emission and
propose the Suzaku sources as plausible counterparts to the
TeV source, although further observations are necessary
to confirm this.
\end{abstract}

\section{Introduction}

The origin and acceleration mechanism of cosmic rays
has remained a long-standing problem since their
discovery nearly 100 years ago.
The most widely accepted acceleration mechanism,
at least for  cosmic rays up to $10^{15.5}$~eV (the knee energy),
is diffusive shock acceleration at the rims of young supernova 
remnants (SNRs)
(e.g., \cite{bell1978,blandford1978}). This hypothesis was
greatly strengthened by the discovery of localized
regions of synchrotron X-ray emission 
from the rims of SNR SN~1006 \citep{koyama1995}, which 
indicated that SNRs were able to accelerate electrons up to $\sim$TeV 
energies. Since then, several other SNRs have been identified as 
electron accelerators
(e.g., \cite{koyama1997,slane2001,bamba2000,bamba2001}) and
numerous studies have indicated that the acceleration process 
is very efficient
(e.g., \cite{vink2003,bamba2003,bamba2005,voelk2005}).
Many papers regard RX~J1713$-$3946
as a proto-type of cosmic-ray accelerating SNRs,
since it is bright in both X-ray and TeV bands \citep{uchiyama2003}.
To date the best evidence for the acceleration of hadronic cosmic 
rays comes from observations of the Tycho SNR \citep{warren2005}.

Recently, 
a TeV $\gamma$-ray telescope, High Energy Stereoscopic System (HESS),
has discovered some new sources in the Galactic plane
\citep{aharonian2005,aharonian2006}.
Some of these sources are extended on scales of $\sim \timeform{0.1D}$,
suggesting Galactic counterparts like
pulsar wind nebulae (PWNe), SNRs, star-forming regions, and so on.
However, their nature is still unclear: 
mainly because many of the new sources
have no counterparts in other wavebands.
They have been sometimes referred to as ``dark particle accelerators''.
Deep follow-up observations are crucial
to elucidate their nature and hard X-ray observations  are a valuable
tool for this, since above a few keV we do not have to contend with 
absorption by the Galactic plane.
The X-ray satellite Suzaku has large effective area and low background
\citep{mitsuda2006},
making it an excellent choice for such follow-up studies.

HESS~J1804$-$216 is one of the brightest 
of the unidentified (unID) TeV sources
\citep{aharonian2005}.
Its size ($\timeform{0.2D}$) and photon index ($2.72\pm0.06$)
makes it one of the largest and softest sources.
\citet{aharonian2006} proposed two
possible counterparts: 
the radio SNR G8.7$-$0.1 (W30) and a young Vela-like pulsar PSR~J1803$-$2137,
although neither is a definitive identification. The HESS source overlaps
only the southwestern portion of G8.7$-$0.1, which
is about $3/4^\circ$ in diameter. 
The distance is estimated to be $\sim$5~kpc based on associating
the SNR with nearby  H~II regions
\citep{kassim1990,finley1994}.
From CO observations, the surrounding region is known
to be associated with molecular gas
where massive star formation takes place \citep{ojeda-may2002}.
From the observed dispersion measure, 
PSR~J1803$-$2137 is located at a distance of $\sim$5.3~kpc
\citep{kassim1990}.
Both candidates were detected by ROSAT \citep{finley1994};
the soft X-ray emission of G8.7$-$0.1 is only apparent along the 
northeastern radio rim.
Very recently, \citet{brogan2006} reported
the discovery of a new SNR (G8.31$-$0.9) in this region.
Swift independently discovered some X-ray sources in this region recently
\citep{landi2006},
but the statistics was quite limited due to the short observation.
It is interesting to note that the
AGASA and SUGAR groups have reported
an excess of EeV cosmic rays from the general direction 
(i.e., within $\sim$20$^\circ$)
toward HESS~J1804$-$216
\citep{hayashida1999,bellido2001},
although this result has not been confirmed by more sensitive
observations by the Pierre Auger Observatory \citep{letessier2005}.
The evidence that HESS~J1804$-$216 is a proton accelerator
therefore remains ambiguous \citep{fatuzzo2006}.

In this paper, we report the first results of
Suzaku follow-up observations of HESS~1804$-$216.
\S\ref{sec:obs} introduces the observations, our analysis
is explained in \S\ref{sec:analysis} and we
discuss our findings in \S\ref{sec:discuss}.

\section{Observations and Data Reduction}
\label{sec:obs}

We observed a source and background region
with Suzaku \citep{mitsuda2006},
which has two sets
of active instruments: four X-ray Imaging Spectrometers (XIS,
\cite{koyama2006}) each at the focus of an X-Ray Telescope (XRTs,
\cite{serlemitsos2006}) and a separate Hard X-ray Detector 
\citep{takahashi2006,kokubun2006}.  Here, we present the
analysis of XIS data only. For this study we observed both a source region
(centered on the published position of HESS~J1804$-$216 and covering the
brightest part of the TeV source) and a
background region some 30$^\prime$ to the southwest that avoided the
TeV source (see  Figure~\ref{HESS-image}).
Observational details are given in  Table~\ref{obslog}.

The XIS was operated in normal full-frame clocking mode
in both pointings.
The data reduction and analysis were made
using HEADAS software version 6.0.6. 
We used version 0.7 processed Suzaku data,
which is an internal processing applied to the Suzaku data 
during the science working group phase
for the purpose of establishing the detector calibration 
as quickly as possible (for the detail, see e.g., \cite{fujimoto2006}).
The XIS pulse height data for each X-ray event were converted to
pulse invariant channels
using the {\tt xispi} software version 2005-12-26
and charge transfer inefficiency parameters from 2006-05-24.
We filtered out data obtained during passages through the 
South Atlantic Anomaly
or with elevation angle to the Earth's limb below  $\timeform{5D}$.
Events with ASCA grades of 0, 2, 3, 4, and 6 were retained.
The total available exposures are
$\sim$40~ks and $\sim$43~ks for source and background,
respectively.

\section{Results}
\label{sec:analysis}

\subsection{Images}

Figure~\ref{xis_image} shows the XIS images
in the 0.5--2.0~keV (left) and 2.0--7.0~keV (right) bands.
All four XIS images were combined, corrected for vignetting,
and smoothed with a $\sim$20$^{\prime\prime}$ (sigma) Gaussian.
In the soft X-ray band, we see no significant source
at the best-fit position of HESS~J1804$-$216
(large cross).
On the other hand, there are two discrete sources 
in the hard band image (the right panel): at
positions (epoch J2000) of
$(\timeform{18h04m44s}, \timeform{-21d42m42s})$
and $(\timeform{18h04m34s}, \timeform{-21d40m20s})$.
We designate them as
Suzaku~J1804$-$2142 and Suzaku~J1804$-$2140.
Note that the systematic position uncertainty is about 1~arcmin.
Hereafter, we will refer to these as Src~1 and Src~2, respectively.
In order to examine their significance,
we extracted source photons
from a circle of 1.5~arcmin radius for Src~1
and 1.8~arcmin radius for Src~2 (see circles in right panel 
of Figure~\ref{xis_image}).
The total 2.0--7.0~keV count rate was
3.7$\times 10^{-2}$~cnt~s$^{-1}$ and 
5.3$\times 10^{-2}$~cnt~s$^{-1}$,
respectively.
The background counts in the same energy range was estimated from
the background observation as 2.8$\times 10^{-3}$~cnts~s$^{-1}$arcmin$^{-2}$.
The significance of Src~1 and Src~2 are, then,
12.4$\sigma$ and 14.9$\sigma$, respectively.
The Galactic ridge emission is known to have spatial fluctuations
of about 30\% in intensity
\citep{yamauchi1996,kaneda1997}.
Assuming a background rate that is 30\% higher than our nominal
one, does not change the source detection significances critically.
Data obtained at low geomagnetic cut-off rigidity (COR) values tends to have
higher background,
so different ranges of COR values in the source and background
pointings could introduce a difference in background level.
We checked the importance of this by heavily filtering on COR
(allowing only values greater than 10~GV) 
and found that we could still detect the sources significantly
even after such a drastic reduction of the data.

The spatial extent of the sources were also examined.
The half power diameter of the XRT is about 2~arcmin
and is almost independent of X-ray energy \citep{serlemitsos2006}.
We compared the counts within the circle of 1~arcmin radius
with the counts from within a radius of 1.5~arcmin (Src~1) and 
1.8~arcmin (Src~2).  According
to XRT calibration data the ratio of counts from a point source
for these radii should be 71\% and 63\%.
The background estimation was done in the same way as discussed above.
We found that
about 75\% (Src~1) and 42\% (Src~2) of the counts are concentrated 
within the 1~arcmin-radius circle.
Therefore, we conclude that
Src~1 is point-like or compact
compared to the spatial resolution of Suzaku,
whereas Src~2 is significantly extended or multiple.

Src~2 is spatially closest to the published
position of HESS~J1804$-$216
(see Figure~\ref{xis_image}), while both Src~1 and Src~2
are within the brightest parts of the TeV source.
No counterparts were found in the SIMBAD database, catalogs of
H~II regions \citep{ojeda-may2002,kassim1990}
and  ultra-compact H~II regions
\citep{bronfman1996}.
Figure~\ref{HESS-image} shows the
TeV $\gamma$-ray image of HESS~J1804$-$216 (grayscale) overlaid with
contours of the XIS 2.0--7.0~keV (white) and soft X-ray 
(ROSAT) emission (black).
Both ROSAT and ASCA imaged the Suzaku XIS 
field of view, but failed to detect emission from either Src~1 or 2,
due to low detection efficiency (ROSAT) in the hard X-ray band
and strong stray light from a nearby bright source (ASCA).
the Suzaku sources are quite compact compared to the HESS source.
Note that  PSR~1803$-$2137 and the previously detected soft X-ray
emission from the northeastern shell of G8.7$-$0.1 fell outside
the XIS field of view (Figure~\ref{HESS-image}).

We extracted light curves for the new X-ray sources,
rejecting data obtained in low COR regions ($<$10~GV),
and applied a Kolmogorov-Smirnov test for time variability.
No significant variability was detected for both sources
during the observation,
with the probability of constancy of 0.38 (Src~1) and 0.49 (Src~2).

\subsection{Spectra}

Source spectra were extracted from within circular regions
of radius of 1.5~arcmin (Src~1) and 1.8~arcmin (Src~2).
Background spectra were obtained from the background observation 
in order to account for the local Galactic ridge emission.
The background spectrum was accumulated from within 
a 3~arcmin radius circle centered on the aim-point of this second
pointing.
The background rate in different observations can
differ due to differences in the average COR.
However our observations were done within one day
of each other 
and the attitude of the satellite was very nearly the same.
The average COR differs by only a few percent between the
source and background observations, so we  have ignored 
any effect of the COR range on the background.

As shown in Figure~\ref{xis_spectra},
the background-subtracted spectra are highly absorbed and featureless,
so we fitted them with absorbed power-law spectral models.
The absorption model used the cross sections of
\citet{morrison1983} with solar abundance \citep{anders1989}.
We made auxiliary files for the effective area using 
{\tt xissimarfgen} \citep{ishisaki2006},
which includes the effect of contamination \citep{koyama2006}.
Figure~\ref{xis_spectra} and Table~\ref{xis_spec_para} shows
our best-fit models and parameters within 90\% errors.
In each case the best-fit
result was acceptable statistically
with reduced $\chi^2$ values of 47.1/46 for Src~1 and 34.6/38 for Src~2.
It was also possible to obtain acceptable fits using a non-equilibrium 
ionization thermal plasma model \citep{borkowski2001}; however the
derived temperatures, 80 ($>$21)~keV for Src~1 and 14 ($>$2)~keV for Src~2,
were unrealistically high.  We conclude that a thermal interpretation for
these spectra is unlikely and favor a nonthermal interpretation instead.

Given the large mis-match in sizes between the new X-ray sources and
HESS~J1804$-$216 we also investigated how much diffuse hard X-ray
emission there could be extending over the entire XIS field of view.
We extracted photons from the entire field of the source pointing,
excluding Src~1 and Src~2. A background spectrum was made using all
the photons in the background pointing.  There was no significant
excess in the source pointing compared to the background over the
2.0--10.0~keV band.  Assuming a power-law model with fixed photon
index $(\Gamma = 2)$ and high absorption ($N_H = 10^{23}$~cm$^{-2}$),
we obtain an absorbed flux upper limit (within a 90\% error) of
$5\times 10^{-13}$~ergs~s$^{-1}$~cm$^{-2}$.

\section{Discussion}
\label{sec:discuss}

The new hard X-ray sources we have successfully detected with Suzaku
are now the nearest potential counterparts
to HESS~J1804$-$216. Although the morphology of TeV $\gamma$-ray emission 
and hard X-rays are quite different, this is also the situation for
the radio SNR G8.7$-$0.1 and the soft X-ray emission.  None of the possible
counterparts is a morphological match to the HESS source. Nevertheless
the new X-ray sources provide interesting new information on the nature
of HESS~J1804$-$216 and the environment.  
An independent report \citep{landi2006} notices new X-ray sources
in this region using Swift.
The position and the flux of the source nearest to HESS~J1804$-$216 is
consistent with those of Src~2,
indicating that we successfully confirmed their detection.
We estimated the chance coincidence probability of unrelated X-ray sources
for HESS~J1804$-$216.
With the $\log N$-$\log S$ relation of X-ray sources in the 2.0--10.0~keV band
\citep{sugizaki2001,yamauchi2002,ebisawa2005},
we found that the expected number of X-ray sources is
only 4--9$\times 10^{-3}$ in the elliptical region
with radii of 0.016 and 0.018 degrees
(the position uncertainty of the HESS source; \cite{aharonian2006}).
Therefore, we concluded that they are physically associated with 
HESS~J1804$-$216.
In the following we discuss
Src~1 and Src~2 separately.

The best-fit absorbing column for Src~1,  
$0.2_{-0.2}^{+2.0}\times 10^{22}$~cm$^{-2}$,
is consistent with the Galactic neutral hydrogen column
in this direction of 
$1.5\times 10^{22}$~cm$^{-2}$ \citep{dickey1990}.
The unabsorbed luminosity is 
$7.5\times 10^{32}(d/{\rm 5~kpc})^2$~ergs~s$^{-1}$
in the 2.0--10.0~keV band,
where $d$ is the source distance.  We quote results for a nominal
value of 5 kpc although in fact we have no independent  constraints on 
the distances to the new X-ray sources.
Src~1 has a very flat spectrum, flatter than typical for
young pulsars and their associated wind nebulae (PWN)
(typical $\Gamma$ values of 1.5 to 2.5; \cite{possenti2002}) 
or shell-like SNRs emitting synchrotron X-rays
(typical $\Gamma$ values of 2.3 to 3).
\citet{uchiyama2002} have reported a region near SNR G78.2+2.1
(Gamma-Cygni) with a photon index of 0.8--1.5, that might be
comparable to Src~1. High mass X-ray binaries also have hard
photon indices ($\Gamma$ of 0.5--2.5; e.g. \cite{muno2003}).
So Src~1 might be this type of source.

In contrast Src~2 has a best-fit absorbing column that
is about an order-of-magnitude higher than the 
expected Galactic column. This is consistent with 
Src~2 being embedded in dense gas, although there
are no cataloged molecular clouds in this direction.
The X-ray and radio counterpart to HESS~J1813$-$178
\citep{brogan2005} also appears to show evidence for
excess absorption.  
The unabsorbed 2--10 keV band luminosity of Src~2,
$1.3\times 10^{33}(d/{\rm 5~kpc})^2$~ergs~s$^{-1}$,
best fit photon index (see Table~\ref{xis_spec_para}),
and extended nature are all consistent with either 
SNR origin: a PWN or shell-like synchrotron emission.

Their origin is still unknown,
but there is some interesting scenario for their nature;
old SNRs colliding with molecular clouds.
\citet{yamazaki2006} argue that an
old SNR shock (age of $10^5$ yr) interacting with a
giant molecular cloud
can emit hard nonthermal X-rays 
and strong TeV $\gamma$-rays from shock accelerated protons 
through $\pi^0$ decay (see also {\cite{fatuzzo2006}).
They can explain large TeV $\gamma$-ray to X-ray
flux ratios (factors of $\sim$100).
Src~2 might represent the X-ray component to the HESS source
under this scenario, with the large excess absorption as a
key piece of evidence.

Comparing images in other wavelengths also suggests the SNR origin.
Soft X-ray emission and HESS source are
both encompassed by the radio contours of G8.7$-$0.1,
suggesting a common origin,
although they do not overlap,
On the other hand, the heavily absorbed Suzaku-discovered 
hard X-ray sources
are located right at the position of peak HESS emission.
The anti-correlation of HESS and ROSAT emission 
is likely a result of a dense region near the SNR which
absorbs the soft X-rays, while at the same time enhancing the production
of TeV emission from $\pi^0$ decay.
The larger size of the TeV source compared to the hard X-ray source
might be due to the much larger diffusion length of protons 
(which provide the TeV $\gamma$-rays) compared to primary electrons 
(which emit the hard X-rays).

HESS~J1804$-$216 has
the large ratio of TeV $\gamma$-ray to hard X-ray fluxes,
which should be an important key to solve its nature.
Table~\ref{ratio} summarizes
X-ray and TeV $\gamma$-rays fluxes for a sample
of TeV $\gamma$-ray emitters.
Generally speaking,
PWNe have rather small values of $F_{TeV}/F_X$,
whereas SNRs tend to be brighter in TeV $\gamma$-rays.
For HESS~J1804$-$216, we assumed three counterparts;
Src~1, 2, and the upper limit of the diffuse emission.
We extracted the diffuse emission from the entire region of XIS FOV,
whereas the TeV emission comes from wider region
($\sigma$ = 0.2 degree; \cite{aharonian2006}).
Therefore, we re-normalized the flux by assuming that
X-ray comes uniformly from the same region of TeV $\gamma$-rays.
The flux in the TeV band is the integrated one
from the entire region of the HESS source.
The ratio of HESS~J1804$-$216 is significantly larger than
that of PWNe \citep{willingale2001,aharonian2004a},
and known shell-like SNRs emitting synchrotron X-rays
\citep[for example]{ozaki1998,aharonian2005c}.
In fact, the $F_{TeV}/F_X$ value is the largest 
among the identified objects,
even if we include the upper-limit to the total hard 
X-ray diffuse emission from the entire XIS field of view
The value is consistent with
that of HESS~J1303$-$631,
one of the new class of  ``dark particle accelerators''
(\cite{mukhergee2005,aharonian2005e}, 
see Table~\ref{ratio}).

Under the assumption that the TeV $\gamma$-rays are emitted via 
inverse Compton on relativistic electrons, higher values of
$F_{TeV}/F_X$ suggest weaker magnetic fields, all other things
being equal.
Therefore, the large $F_{TeV}/F_{X}$ of HESS~J1804$-$216 suggest that
the magnetic field in this source is weaker than typical in
traditional PWNe and SNRs,
or inverse Compton origin is unlikely.

\section{Summary}

Thanks to the large effective area and low background of 
the XIS onboard Suzaku, we have
discovered two discrete X-ray sources
positionally coincide with a HESS unidentified source,
HESS~J1804$-$216.
The new X-ray sources have no counterpart in other wavebands,
except for the TeV source.
Their X-ray spectra are hard and deeply absorbed.  The
absorption of Src~2 significantly exceeds the Galactic value
implying that it is deeply embedded in dense gas, which
gives strong support to the idea that the TeV $\gamma$-rays
arise from $\pi^0$ decay.
Additionally the $F_{TeV}/F_X$ value is the largest among known 
sources.
HESS~J1804$-$216 an important object for trying to
untangle the nature of a ``dark particle accelerator.''

Much remains to be done to confirm the nature of these new sources,
especially in the radio band, which should be able to address 
the SNR origin.
X-ray and TeV $\gamma$-ray observations with better spatial resolution
are also needed and high time resolution will help
to judge whether there is a pulsar.

\section{Acknowledgements}

We thank W.~Hoffman, S.~Funk, N.~White, and P.~Ubertini
for fruitful discussions.
We also thank all members of the Suzaku team, with 
our particular thanks to N.~Yamasaki
and the Suzaku operation team who kindly scheduled these observations.
This research has made use of the SIMBAD database,
operated at the CDS, Strasbourg, France. 
This work is supported in part by the Grant-in-Aid for Young Scientists (B) of
the Ministry of Education, Culture, Sports, Science and Technology 
(No.~17740183). JPH acknowledges support from NASA grant NNG05GP87G.

\begin{figure}
\FigureFile(100mm,50mm){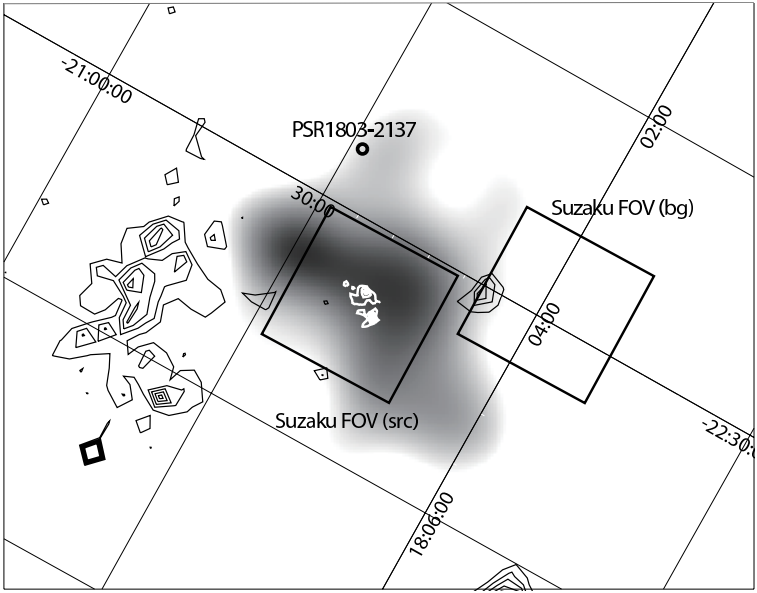}
\caption{HESS image (gray scale; courtesy of W. Hoffman and S. Funk)
overlaid XIS 2.0--7.0~keV image
(white contour),
with J2000 coordinates.
The black contour and circle represents
ROSAT image of G8.7$-$0.1 \citep{finley1994}
and PSR~1803$-$2137.
All images are in the linear scale.
The black boxes show the field of views of XIS.
}
\label{HESS-image}
\end{figure}

\begin{figure}
\begin{center}
\FigureFile(80mm,40mm){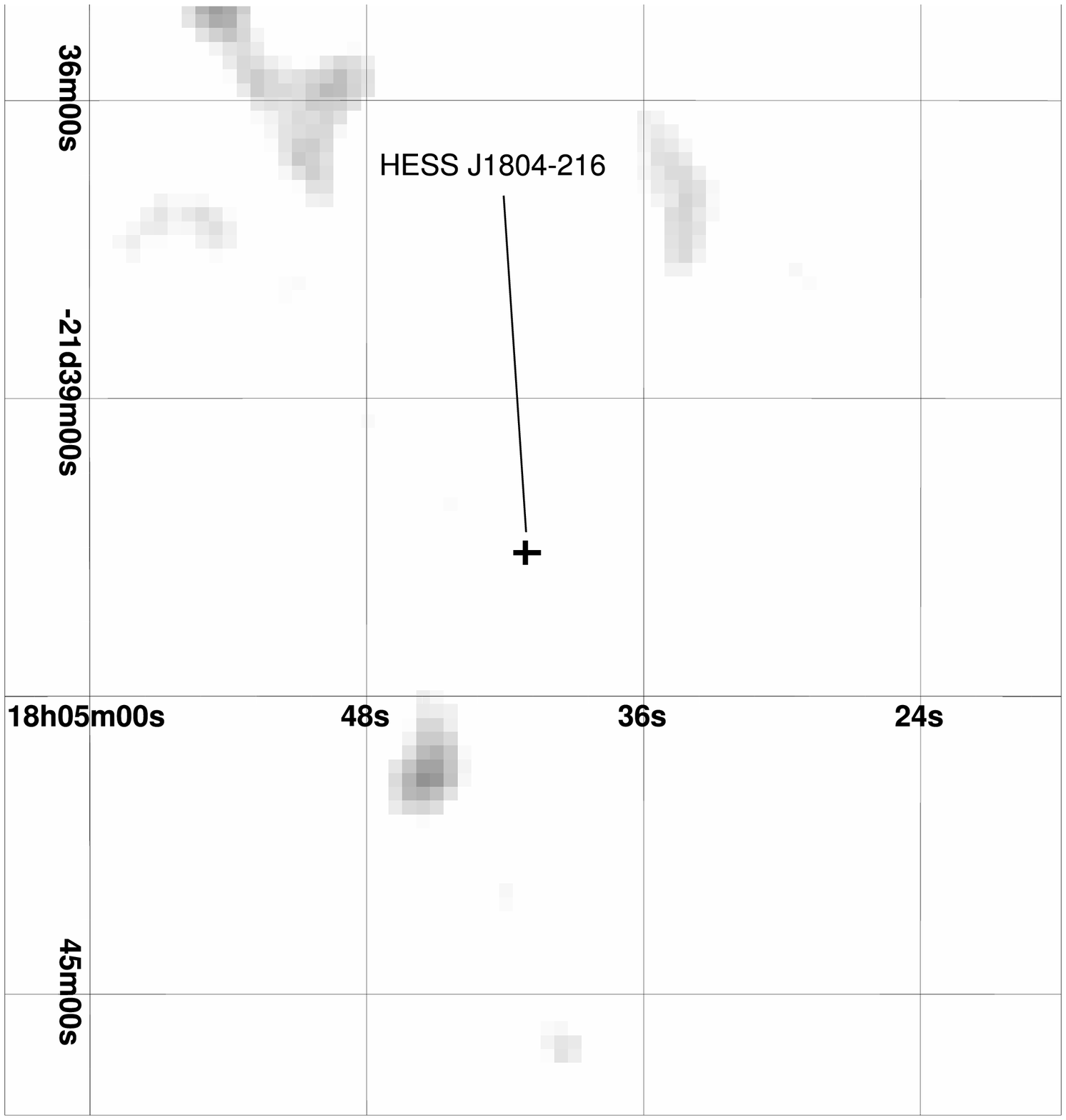}
\FigureFile(80mm,40mm){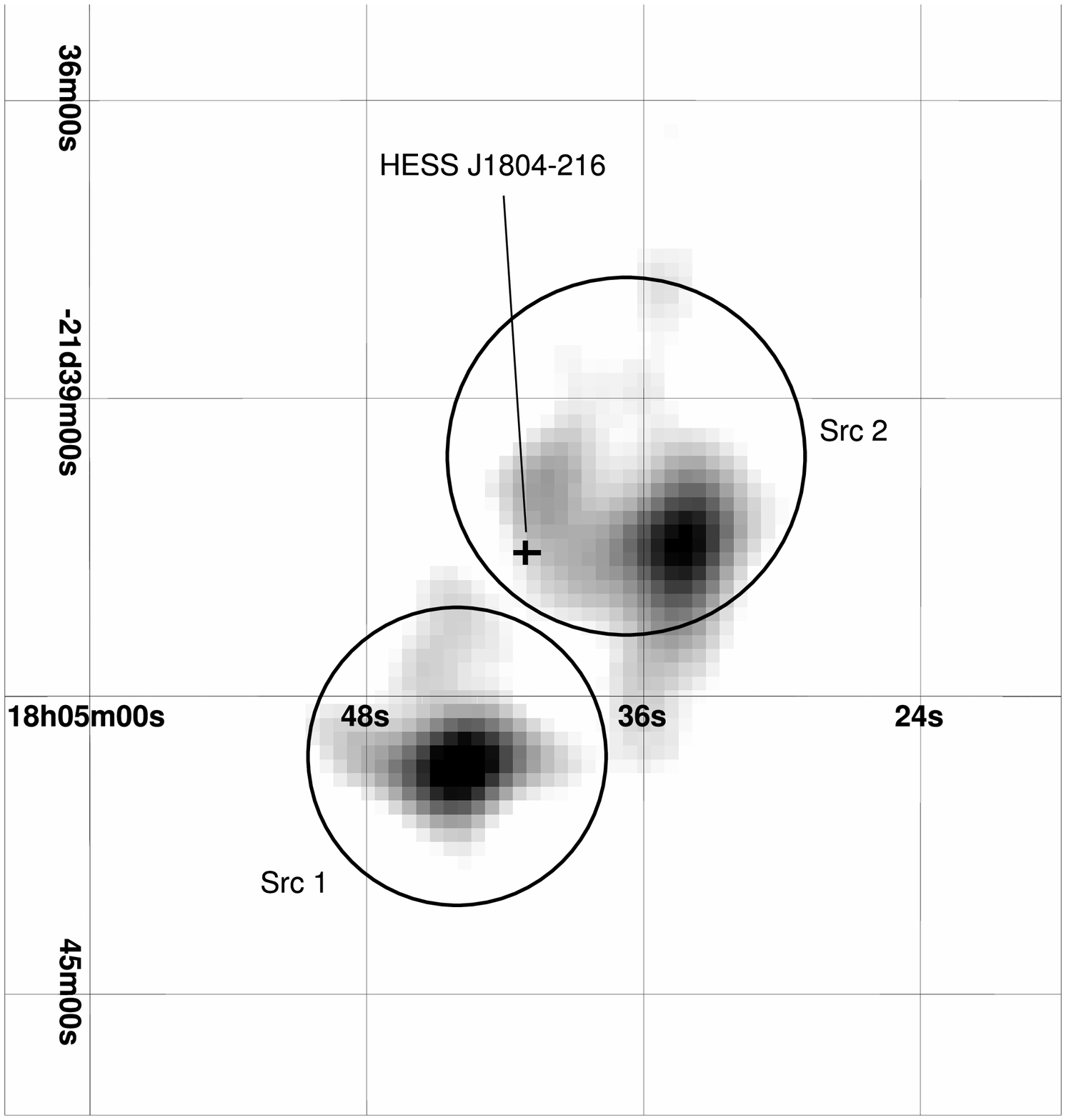}
\end{center}
\caption{XIS images of HESS~J1804$-$216 region
in the 0.5--2.0~keV band (left) and 2.0--7.0~keV band (right),
with J2000 coordinates.
The scales are linear.
Each picture is smoothed with $\sim$20~arcsec scale.
The thick crosses in both panels represents
the best-fit position of HESS~J1804$-$216.
The source regions used for the timing and spectral analysis
are shown by thin solid lines in the right panel.}
\label{xis_image}
\end{figure}

\begin{figure}
\begin{center}
\FigureFile(80mm,50mm){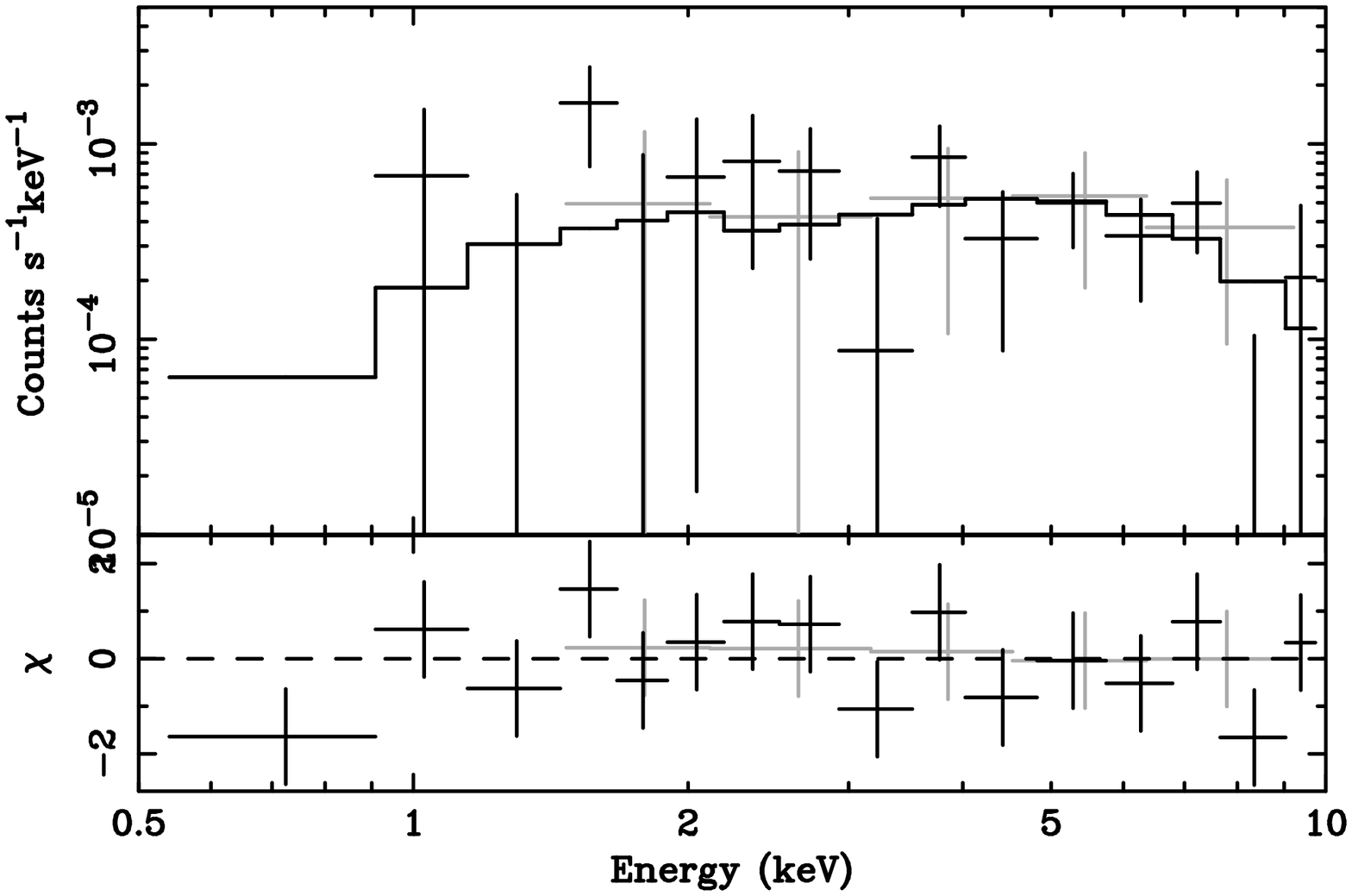}
\FigureFile(80mm,50mm){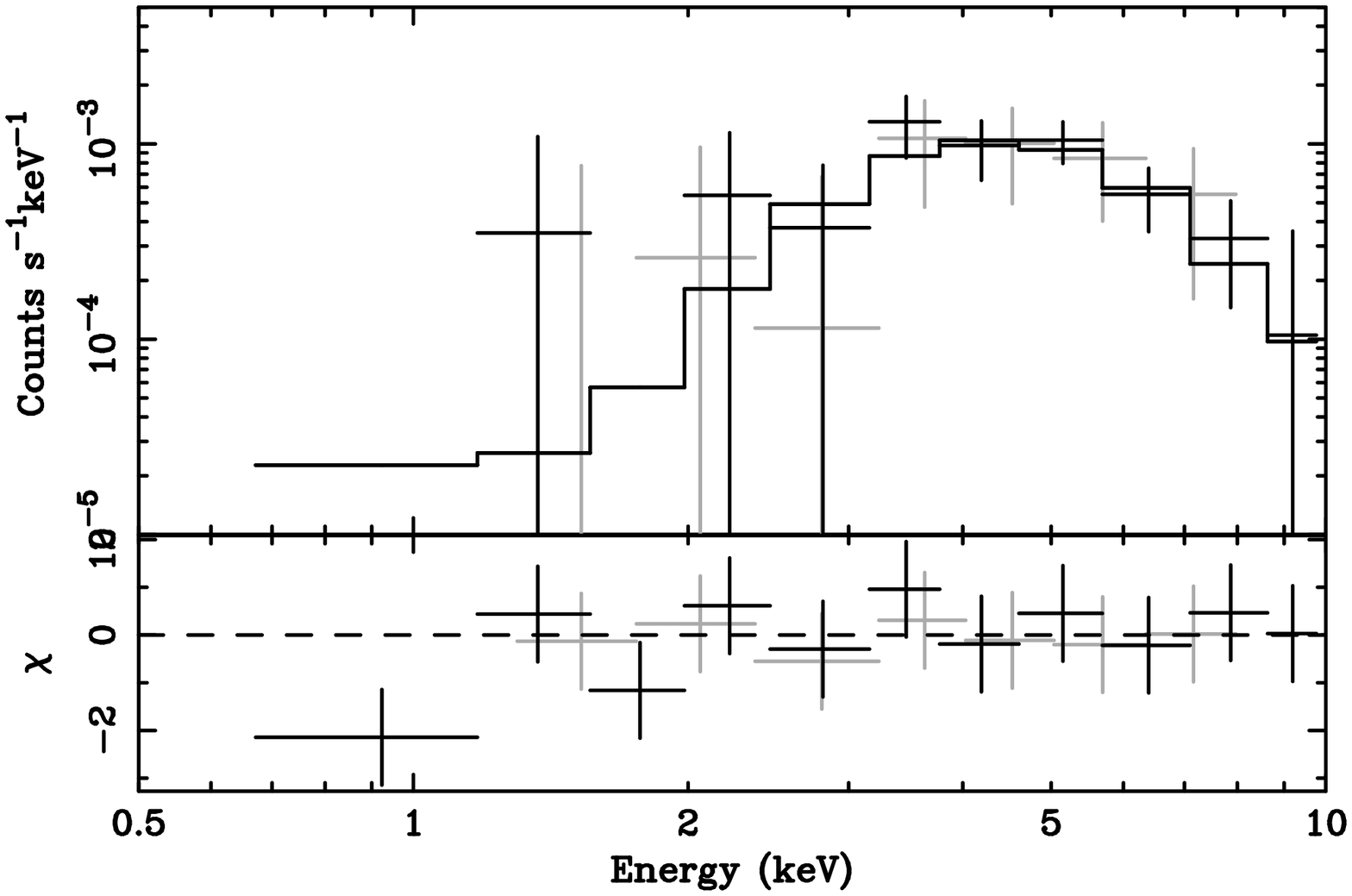}
\end{center}
\caption{Spectra of the Src~1 (left) and Src~2 (right).
Gray and Black crosses represent
XIS0+2+3 (added just for the clear viewing) and XIS1 data, respectively.
The best-fit power-law models are shown with solid lines.
Lower panels in the figures are data residuals from the best-fit models.
}
\label{xis_spectra}
\end{figure}

\begin{table}[hbtp]
\begin{center}
\caption{Observation Log.}
\label{obslog}
\begin{tabular}{p{10pc}cc}\hline\hline
 & source & background \\
\hline
Date (yyyy/mm/dd) & 2006/04/06 & 2006/04/07 \\
Aim point (RA, Dec.) & (271.1713, $-$21.6759) & (270.9600, $-$22.0244) \\
Exposure (ks) & 40 & 43 \\
\hline
\end{tabular}
\end{center}
\end{table}

\begin{table}[hbtp]
\caption{Best-fit parameters of spectral fittings.$^\dagger$}
\label{xis_spec_para}
\begin{center}
\begin{tabular}{p{15pc}cc}\hline\hline
 & Src~1 & Src~2 \\
\hline
$\Gamma$ \dotfill & $-$0.3 ($-$0.8--0.2) & 1.7 (0.7--3.1) \\
$N_{\rm H}$ [$10^{22}$~cm$^{-2}$]\dotfill & 0.2 ($<$2.2) & 11 (5.3--21) \\
intrinsic $F_{\rm 2-10~keV}$$^\ddagger$\dotfill & 2.5 (2.1--2.9)
 & 4.3 (3.2--8.3) \\
observed $F_{\rm 2-10~keV}$$^\ddagger$\dotfill & 2.5 & 2.4 \\
\hline
\multicolumn{3}{p{12cm}}
{$^\dagger$: Parentheses indicate single parameter 90\% confidence regions.}\\
\multicolumn{3}{p{12cm}}
{$^\ddagger$: In the unit of $10^{-13}$ergs~s$^{-1}$cm$^{-2}$.}
\end{tabular}
\end{center}
\end{table}

\begin{table}
\begin{center}
\caption{Flux ratio between X-ray and TeV $\gamma$-ray bands.}\label{ratio}
\begin{tabular}{p{8pc}ccccc}
\hline\hline
 & $F_{X}$$^a$ & $F_{TeV}$$^b$ & $F_{TeV}/F_{X}$ & Type$^c$ & Ref.$^d$ \\
\hline
Crab nebula\dotfill & $2.1\times 10^4$ & 56 & $2.7\times 10^{-3}$ & PWN & (1) (2) \\
Nebula of PSR~1509\dotfill & 32 & 16 & 0.5 & PWN & (3) (4) \\
G0.9+0.1\dotfill & 5.8 & 2.0 & 0.3 & PWN & (5) \\
SN~1006 NE \dotfill & 19 & $<$2.6 & $<$0.1 & SNR & (6) (7) \\
RX~J1713$-$3946\dotfill & 540 & 35 & 0.06 & SNR & (8) (9) \\
HESS~J1813$-$178\dotfill & 7 & 9 & 1.3 & SNR & (10) (11) \\
HESS~J1303$-$631\dotfill & $<$6.4 & 10 & $>$1.6 & unID & (12) (13) \\
\hline
HESS~J1804$-$216\dotfill & 0.2 (Src 1) & 6.5 & 33 & unID & this work, (11) \\
HESS~J1804$-$216\dotfill & 0.4 (Src 2) & 6.5 & 16 & unID & this work, (11) \\
HESS~J1804$-$216\dotfill & $<$0.8 (diffuse) & 6.5 & $>$8 & unID & this work, (11) \\
\hline
\multicolumn{6}{p{12cm}}
{$^a$: Unabsorbed flux in the 2--10~keV band in the unit of 
10$^{-12}$~ergs~s$^{-1}$~cm$^{-2}$.}\\
\multicolumn{6}{p{12cm}}
{$^b$: Unabsorbed flux in the 1--10~TeV band in the unit of 
10$^{-12}$~ergs~s$^{-1}$~cm$^{-2}$.}\\
\multicolumn{6}{p{12cm}}
{$^c$: PWN: Pulsar wind nebulae, SNR: Supernova remnants,
unID: TeV unidentified sources.}\\
\multicolumn{6}{p{12cm}}
{$^d$: (1) \citet{willingale2001};
(2) \citet{aharonian2004a};
(3) \citet{delaney2006};
(4) \citet{aharonian2005b};
(5) \citet{aharonian2005f};
(6) \citet{ozaki1998};
(7) \citet{aharonian2005c};
(8) \citet{aharonian2004b}; 
(9) \citet{slane2001};
(10) \citet{brogan2005};
(11) \citet{aharonian2006};
(12) \citet{aharonian2005e};
(13) \citet{mukhergee2005};
}
\end{tabular}
\end{center}
\end{table}

\end{document}